\documentclass[twoside,a4paper,11pt]{proceedings}
\usepackage{graphicx}
\usepackage{natbib}
\begin{document}
\pagenumbering{arabic}
\pagestyle{myheadings}
\thispagestyle{empty}
\vspace*{-1cm}
%{\flushleft\includegraphics[width=\textwidth,bb=58 650 590 680]{stamp.pdf}}
%{\flushleft\includegraphics[width=\textwidth,viewport=58 650 590 680]{stamp.pdf}}
{\flushleft\includegraphics[width=3cm,viewport=0 -30 200 -20]{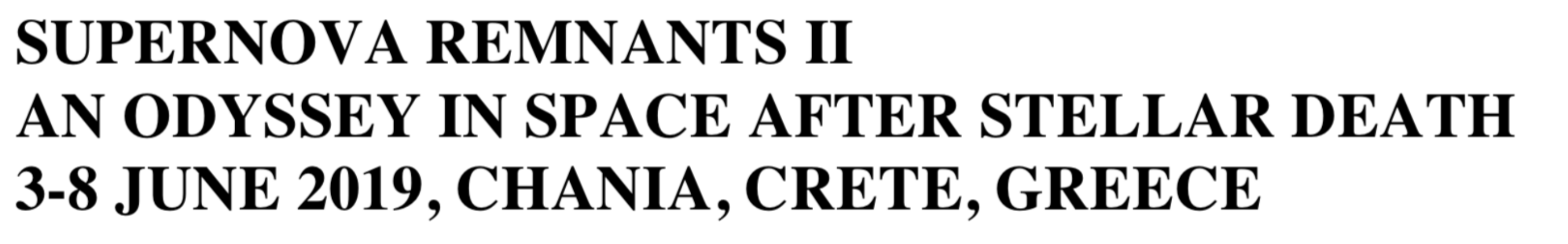}}
\vspace*{0.2cm}
\begin{flushleft}
{\bf {\LARGE
%%% TITLE of the paper.
Evolution of High-energy Particle Distribution in Supernova Remnants
}\\
\vspace*{1cm}
%%% Include here the LIST OF AUTHORS.
%%% Note that the last author has to be preceeded by an AND.
Houdun Zeng$^1$,
Yuliang Xin$^1$,
Qi Fu$^2$
and Siming Liu$^1$
%,
%
% Do not delete next few lines
}\\
\vspace*{0.3cm}
%
%%% AFFILIATIONS LIST.
%%% and the AFFILIATIONS LIST. Note that one affiliation per line.
%%% Add as many affiliations as necessary.
$^{1}$
Key Laboratory of Dark Matter and Space Astronomy,
Purple Mountain Observatory, Chinese Academy of Sciences, Nanjing 210008, China\\
$^{2}$
School of Physical Science and Technology, Lanzhou University, Lanzhou 730000, China
% Do not delete next few lines
\end{flushleft}
% Headings
\markboth{
%%% Type the SHORT version of the paper t
Evolution of High-energy Particle Distribution in SNRs
}{Zeng et al.
}
\thispagestyle{empty}
\vspace*{0.2cm}
\begin{minipage}[l]{0.09\textwidth}
\
\end{minipage}
\begin{minipage}[r]{0.9\textwidth}
\vspace{1cm}
\section*{Abstract}{\small
%%% Type the ABSTRACT of your paper
The spectra fits to a sample of 34 supernova remnants \citep{2019ApJ...874...50Z} are updated.
$\gamma$-ray spectra of 20 supernova remnants (SNRs) with a soft TeV spectrum are
further analyzed. We found that 17 of them can be fitted in the hadronic scenario with a
single power-law ion distribution with an index of $\sim$ 2.6, which is significantly softer than the ion distribution inferred from $\gamma$-ray observations of star-burst
galaxies. If Galactic cosmic rays are mostly produced by SNRs, this result suggests
that SNRs in star-burst galaxies may never reach the phase with a soft $\gamma$-ray
spectra or escape of high-energy particles from SNRs before they reach this phase with
a soft $\gamma$-ray spectrum dominates the contribution of SNRs to Galactic cosmic rays.

\vspace{6mm}
\normalsize}
\end{minipage}
%%% BODY of the paper

\section{Introduction}

As early as 1930s, supernova remnants (SNRs) have been suggested as the dominant contributors to the galactic cosmic rays \citep{1934PNAS...20..259B}.
However, it is still not clear how the galactic cosmic ray spectrum is related to energetic particle distribution in individual SNRs.

\section{Sample and Model}
The sample and model are introduced in Zeng et al. (2019). Note that we here update the gamma-ray data and fitted results of SNR W30 \citep{2019arXiv190700180L} and G205.5+0.5, and remove SNR G150.3+4.5 whose radio emission may be produced by secondary leptons in the
hadronic scenario for the $\gamma$-ray emission.

\section{The Results}

Figure. 1 shows the main results of correlations of model parameters, and the evolution of high-energy particle distribution in SNRs.
There are the detailed introductions for those in \cite{2019ApJ...874...50Z}.

Figure. 2 shows the results of fitted SED of every hadronic original SNR with a single power-law distribution, as well as the fitted index. The fitted index $\Gamma$ and $E_{min}$ are compared with the high-end index ($\alpha +1 $) and break energy ($E_{br}$) of particle distribution in \cite{2019ApJ...874...50Z}.

Figure. 3 and 4 show the results of fitted SED of SNRs and/or Starburst galaxies with different energy bands. Specially, the left panel in Figure. 3 indicates that the reduced $\chi^2$ and index vary with the fitting energy band,
implying that high-energy particles may have been escaped from SNR efficiently so that only particles with relatively low-energies are still trapped.

\begin{figure}
\center
\includegraphics[width=0.3\textwidth]{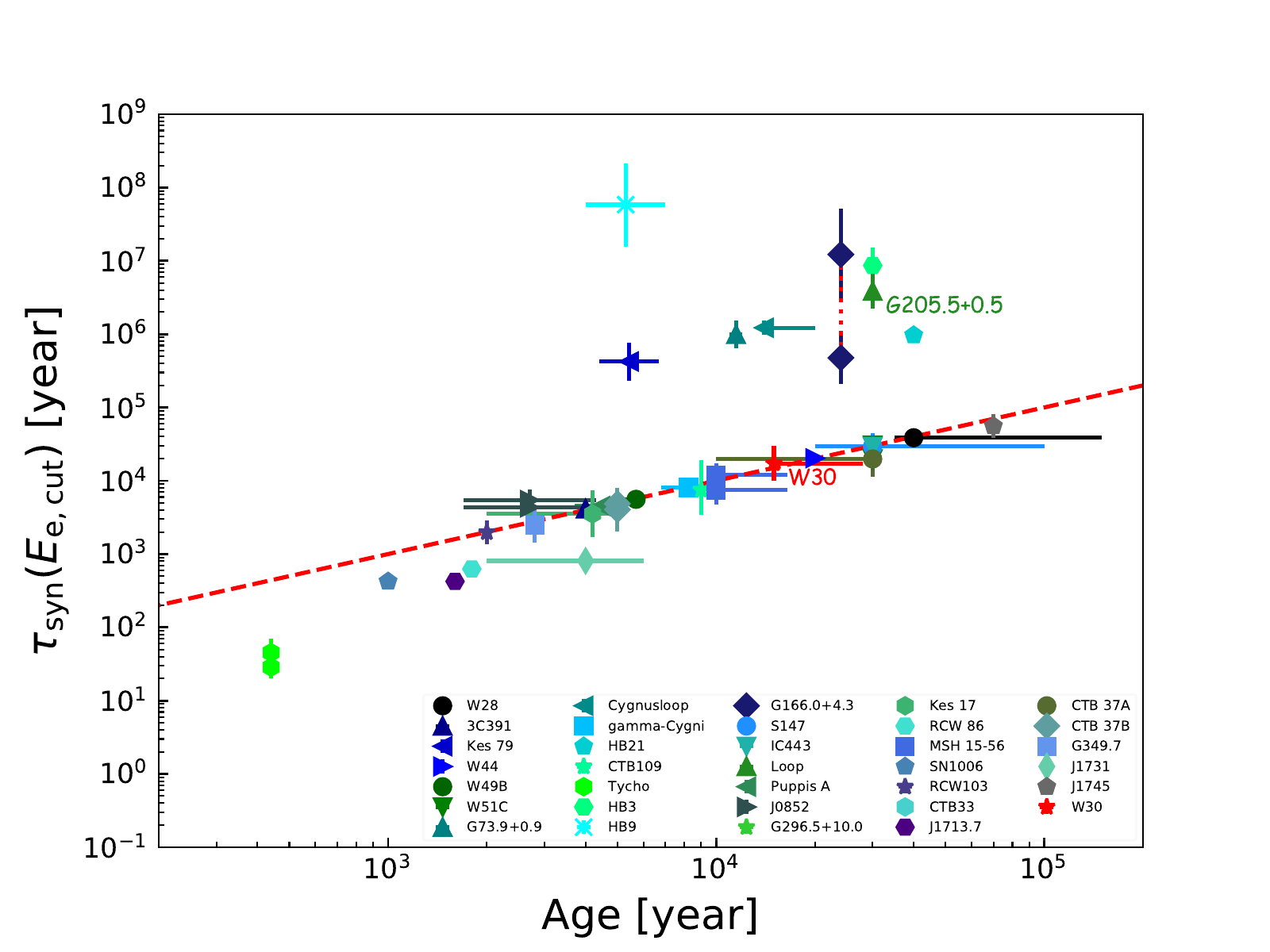}
\includegraphics[width=0.3\textwidth]{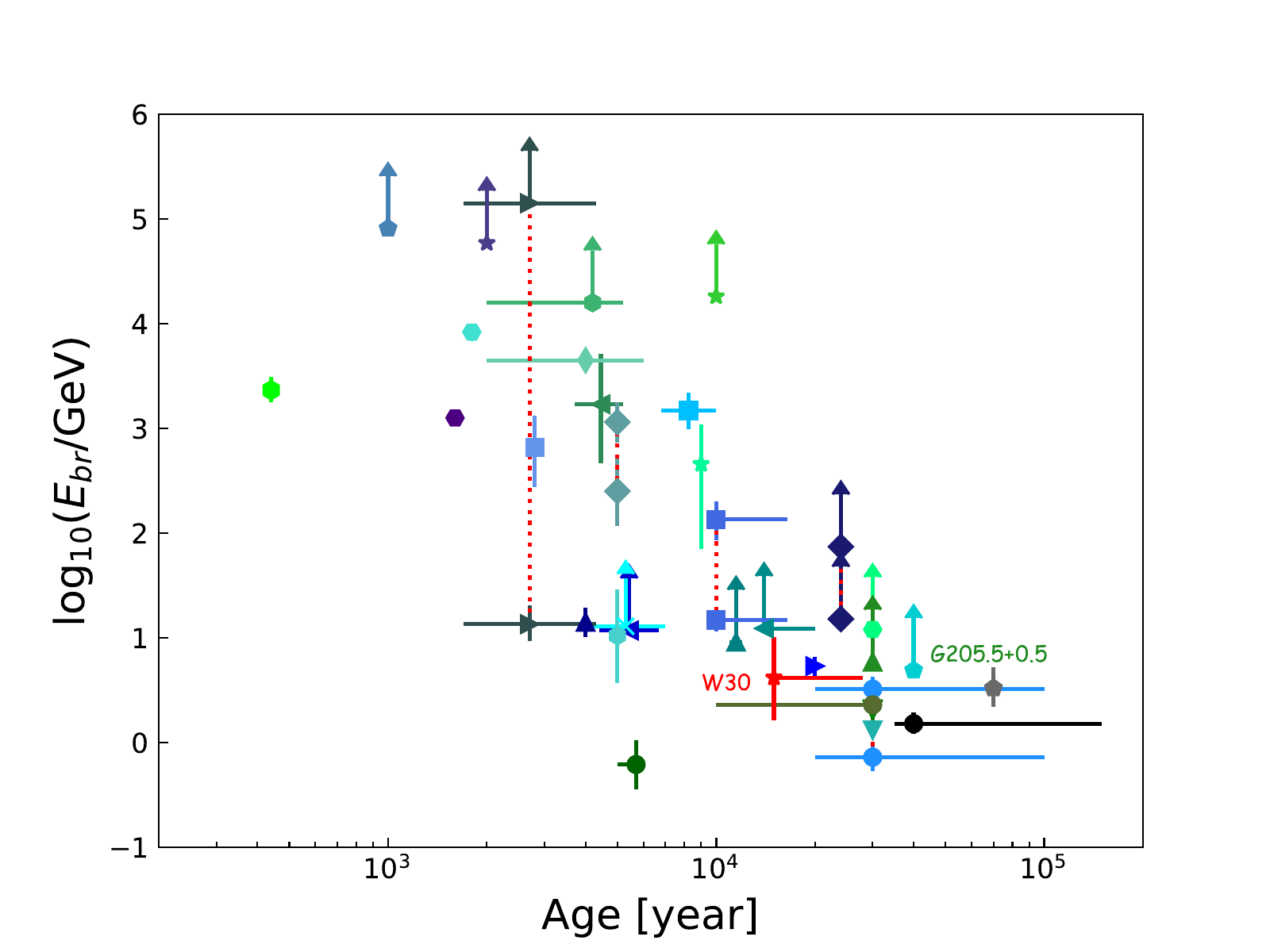} \\
\includegraphics[width=0.5\textwidth]{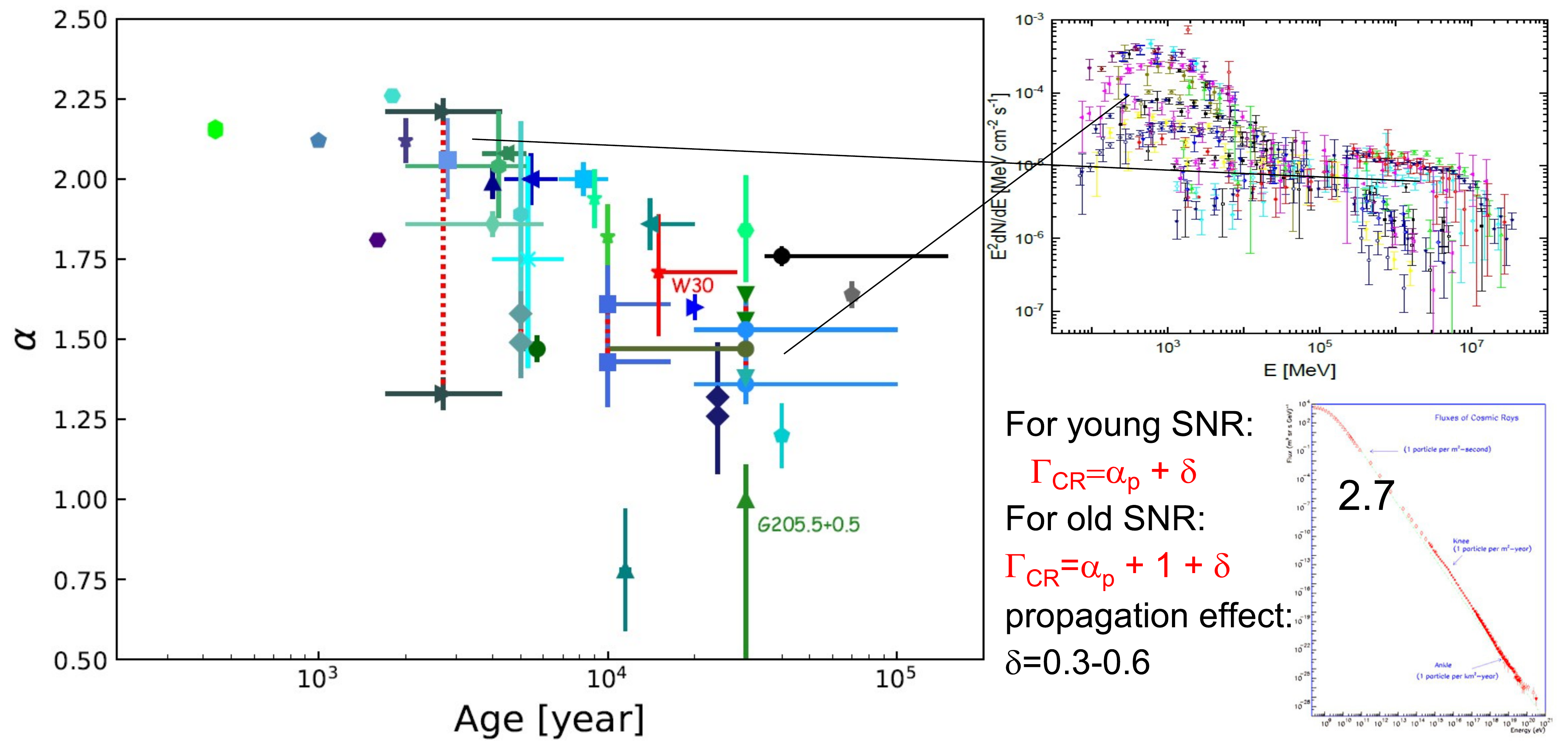}
\caption{Upper: the electron synchrotron cooling time at the cutoff energy and the break energy of particle
distribution decreases with the age of the corresponding SNR. Lower: correlation between low-energy spectral
indices and the ages of SNRs, and the scenario of SNR origin of
Galactic cosmic rays.}
\end{figure}

\begin{figure}
\centering
\includegraphics[width=0.4\textwidth]{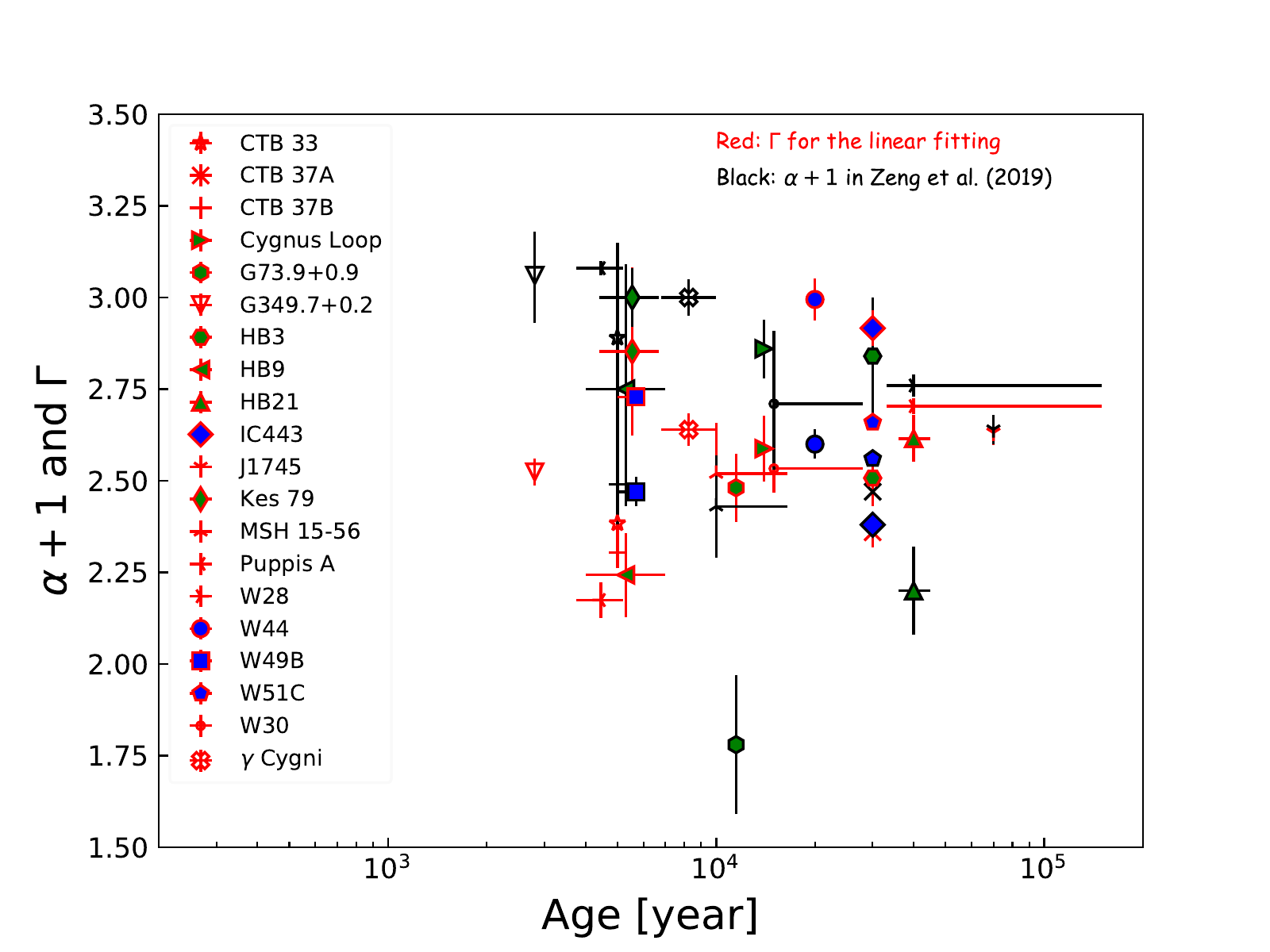}
\includegraphics[width=0.4\textwidth]{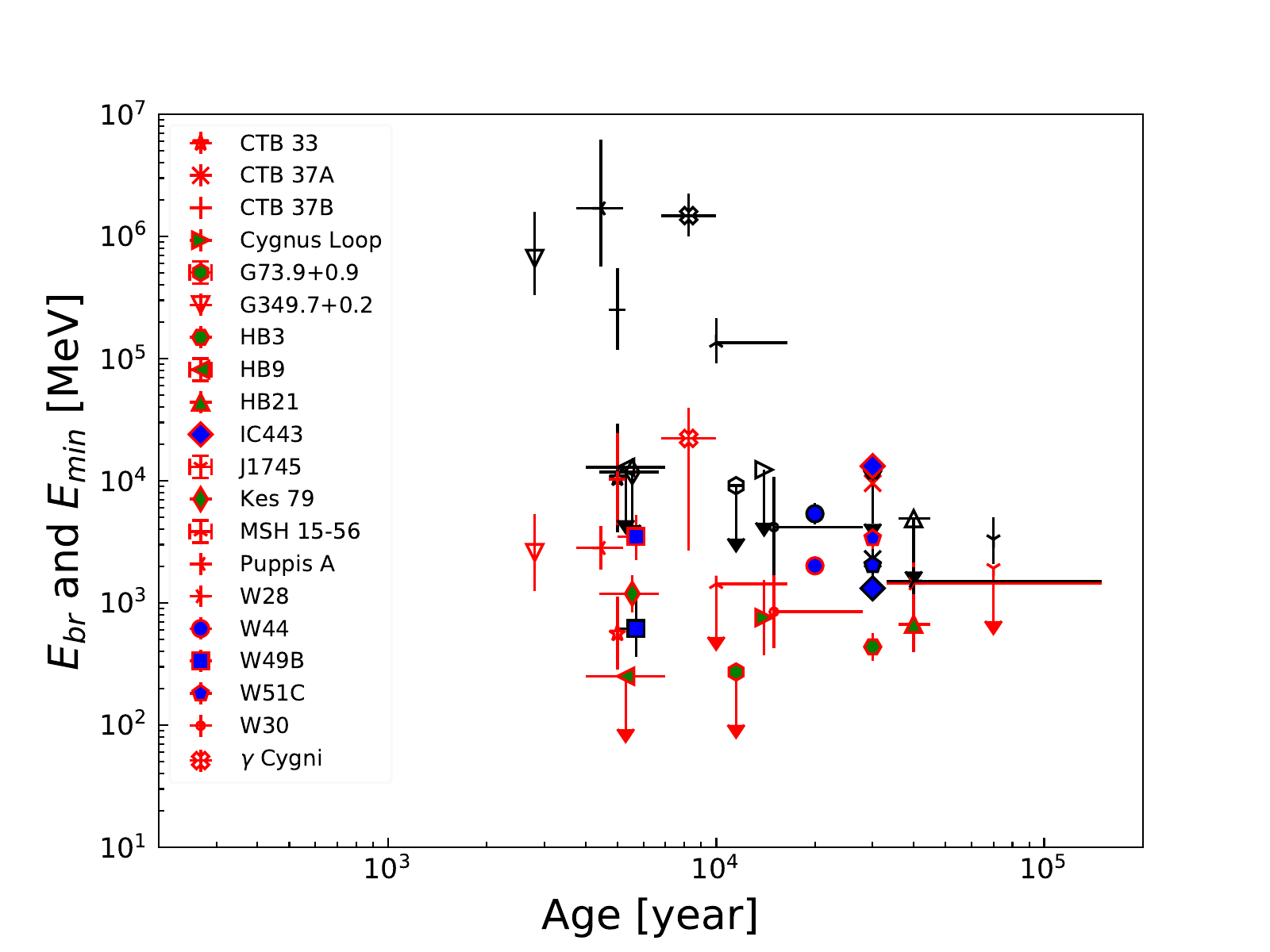}
\caption{Red point: the results of fitted SED of hadronic original SNRs with a single power-law distribution. Here $\alpha$ is the index and $E_{min}$ is the minimun energy of fitted SED. Black point: the values from Zeng et al. (2019).}
\end{figure}

\begin{figure}
\centering
\includegraphics[width=0.32\textwidth]{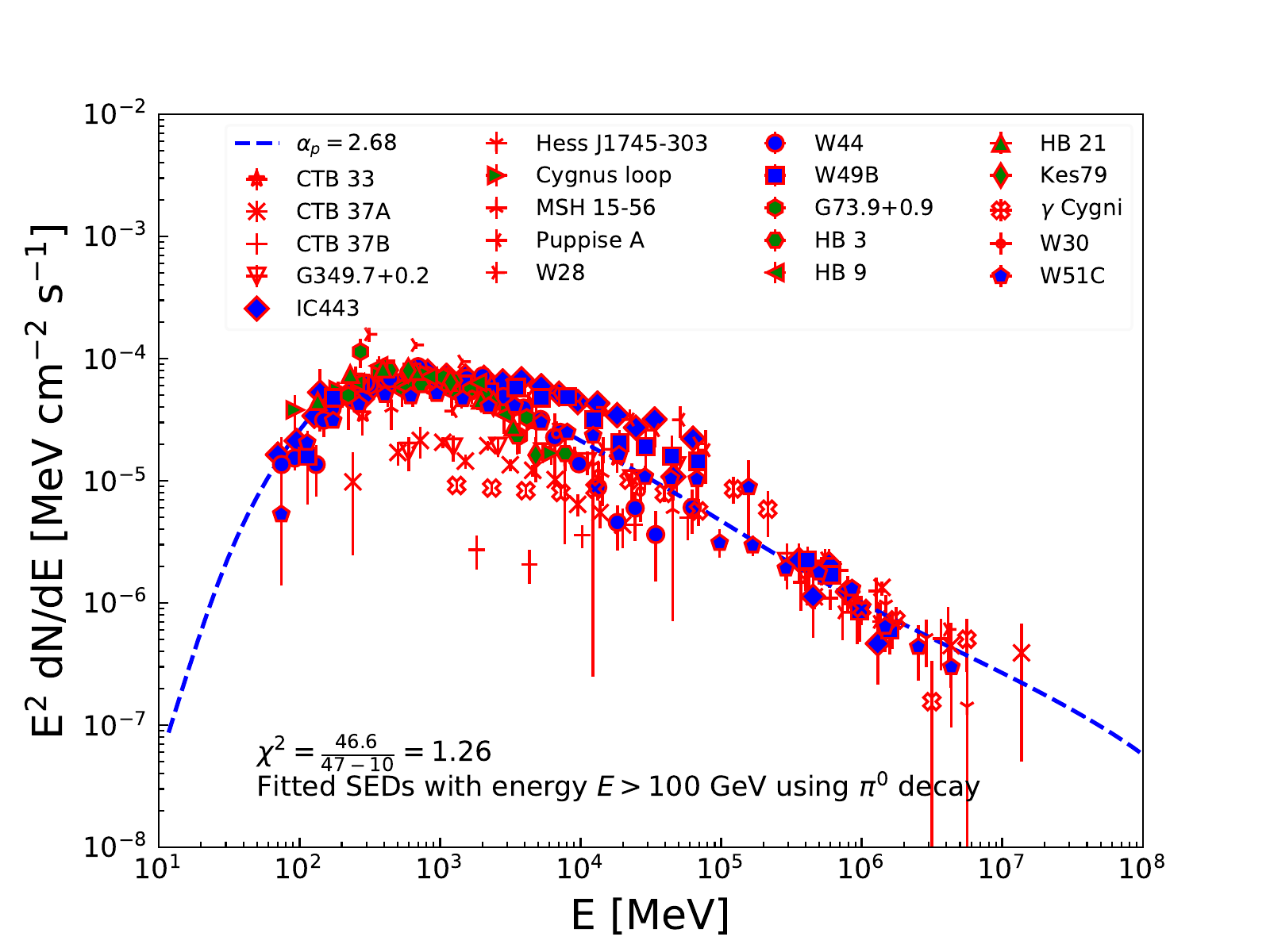}
\includegraphics[width=0.32\textwidth]{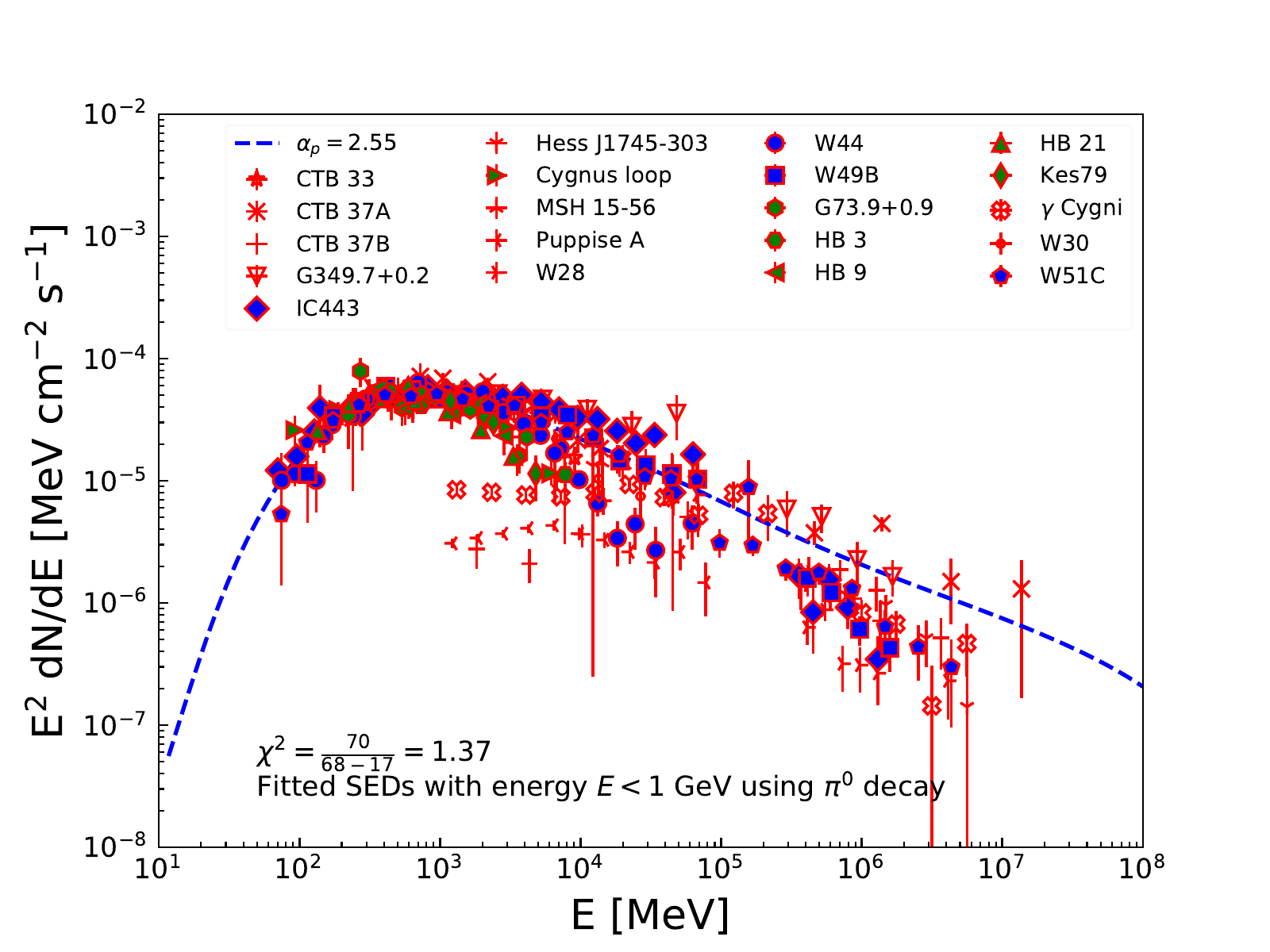}
\includegraphics[width=0.32\textwidth]{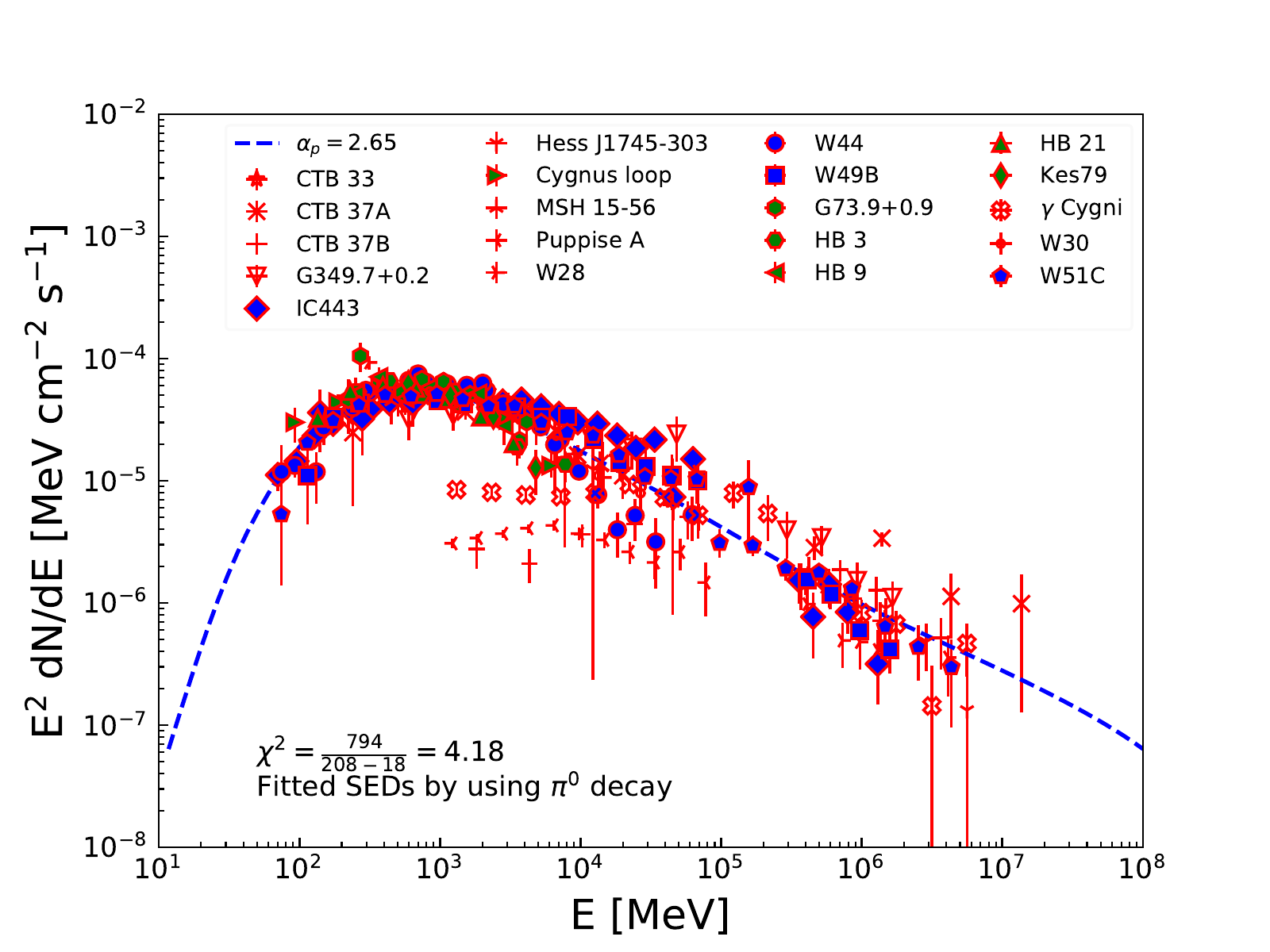}
\caption{Left: the result of fitted SEDs with $E>100$ GeV using the $\pi^{0}$ decay for 9 SNRs. Middle: the result of fitted SEDs with $E<1$ GeV using the $\pi^{0}$ decay for 16 SNRs. Right: the result of fitted SEDs with the whole spectrum using the $\pi^{0}$ decay for 17 SNRs.} %Note that the SNRs with opaque data points  do not participate in %fit. }
\end{figure}

\begin{figure}
\centering
\includegraphics[width=0.4\textwidth]{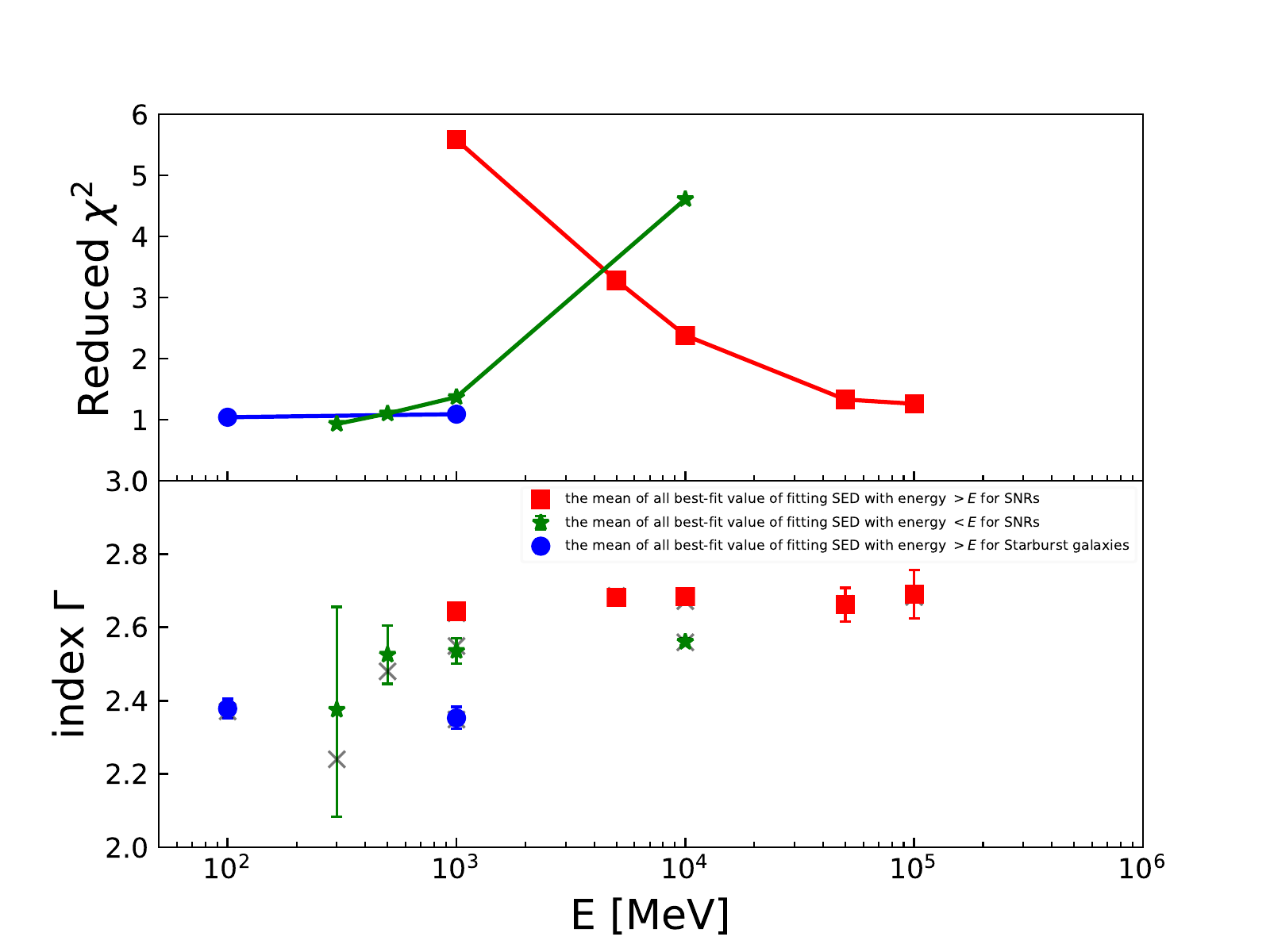}
\includegraphics[width=0.4\textwidth]{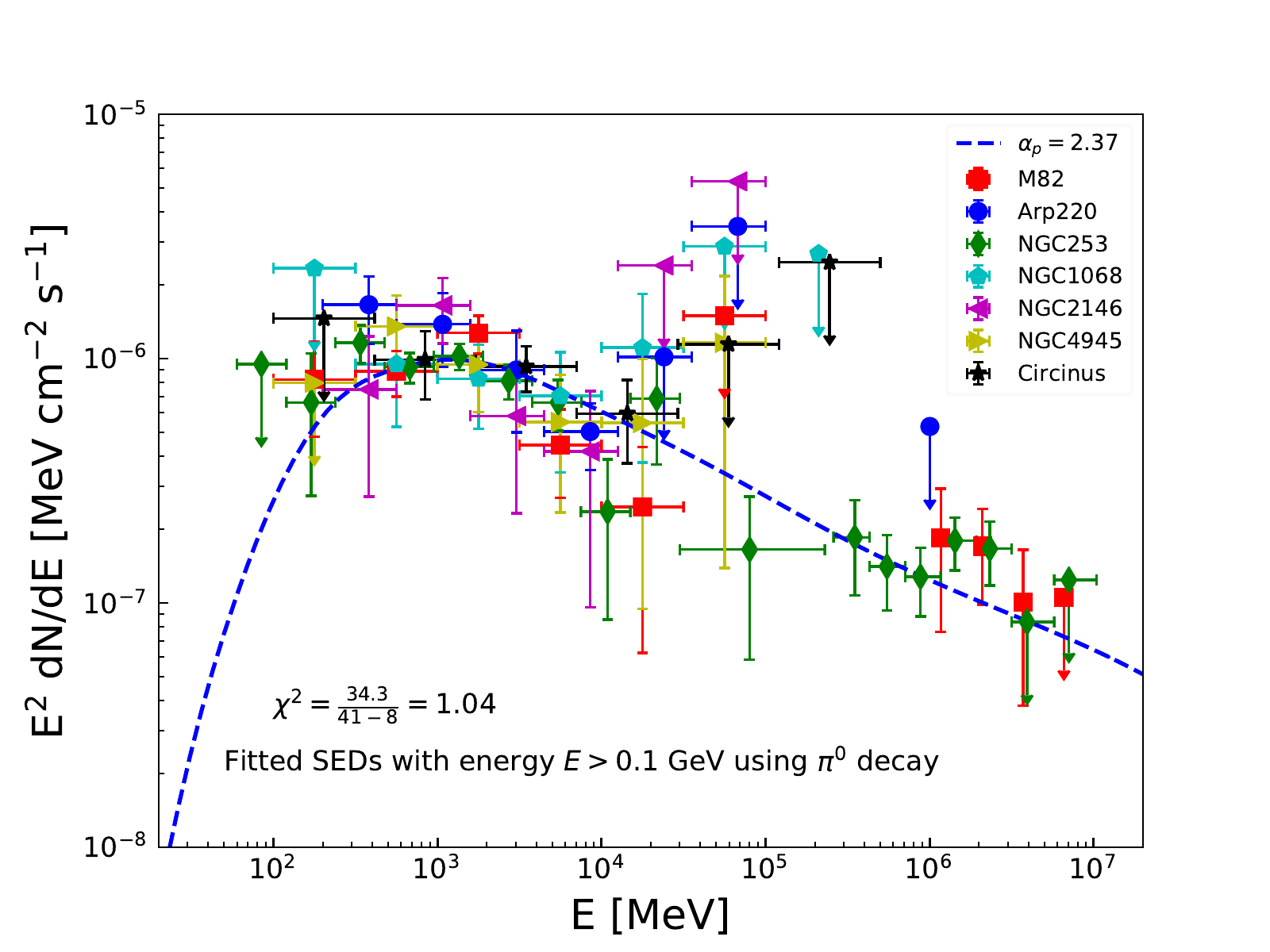}
\caption{Left: the reduced $\chi^2$ and index vary with the fitting energy bands. The red, green and blue points represent the best-fit values of high- and low-end SEDs for SNRs and high-end SEDs for Starburst galaxies. Right: the result of fitted SEDs with $E>0.1$ GeV using the $\pi^{0}$ decay for 7 Starburst galaxies.}
\end{figure}

% Do not delete the next line
\small  % Do not delete
%
%%% Comment the following line if you do not have acknowledgments.
\section*{Acknowledgments}   % Do not delete if you declare acknowledgments
%
%%% ACKNOWLEDGMENTS
%%% ACKNOWLEDGMENTS
Support for this work was provided by National Key R\&D Program
of China: 2018YFA0404203, NSFC grants: U1738122,
11761131007 and by the International Partnership Program of
Chinese Academy of Sciences, grant No. 114332KYSB20170008

%%% BIBLIOGRAPHY
\bibliographystyle{aj}
\small
%\bibliography{proceedings}

\begin{thebibliography}{}

\bibitem[\protect\citeauthoryear{{Baade} \& {Zwicky}}{{Baade} \&
  {Zwicky}}{1934}]{1934PNAS...20..259B}
{Baade}, W.,  \& {Zwicky}, F. 1934, Proceedings of the National Academy of
  Science, 20, 259

\bibitem[\protect\citeauthoryear{{Liu} et~al.}{{Liu}
  et~al.}{2019}]{2019arXiv190700180L}
{Liu}, B., {Yang}, R.-z., {Sun}, X.-n., {Aharonian}, F.,  \& {Chen}, Y. 2019,
  arXiv e-prints, arXiv:1907.00180

\bibitem[\protect\citeauthoryear{{Zeng}, {Xin}, \& {Liu}}{{Zeng}
  et~al.}{2019}]{2019ApJ...874...50Z}
{Zeng}, H., {Xin}, Y.,  \& {Liu}, S. 2019, apj, 874, 50

\end{thebibliography}

\end{document}